# Measurements of the $^{64}$Zn(n,α)$^{61}$Ni Cross Section at $E_n$ =5.0-6.75 MeV

Yuri GLEDENOV[1,*], Milana SEDYSHEVA[1], Pavel SEDYSHEV[1], Alexander OPREA[1]
Zemin CHEN[2], Yingtang CHEN[2], Jing YUAN[2], Guohui ZHANG[3], Guoyou TANG[3]
Gonchigdorj KHUUKHENKHUU[4], Pawel SZALANSKI[5]

[1] *Frank Laboratory of Neutron Physics, JINR, 141980 Dubna, Russia*
[2] *Department of Physics, Tsinghua University, Beijing, China 100084*
[3] *Institute of Heavy Ion Physics, Peking University, Beijing, China 100871*
[4] *Nuclear Research Centre, National University of Mongolia, Ulaanbaatar, Mongolia*
[5] *University of Lodz; High School of National Economy in Kutno, Poland*

The experiment of determination of the $^{64}$Zn(n,α)$^{61}$Ni reaction cross section in the 5.0 - 6.75 MeV neutron energy range was performed at the 4.5 MV Van de Graaf accelerator at the Institute of Heavy Ion Physics, Peking University, Beijing. Double section ionization chamber with grid was used for direct registration of the reaction products. The cross sections and angular distributions were extracted from the experimental data. The obtained values were compared with the results of other authors and theoretical estimations. The analysis of the experimental data and model calculations were carried out.

*KEYWORDS: (n,α) reaction, cross sections, angular distributions, ionization chamber*

## I. Introduction

Investigation of $^{64}$Zn(n,α)$^{61}$Ni reaction in the energy range of several MeV is of great importance in various applied and basic fields. The data for the $^{64}$Zn(n,α)$^{61}$Ni reaction cross section is important for estimation of radiation damage in structural materials of nuclear reactors. On the other hand the information about the cross section and angular distribution is of interest for researching of nuclear reaction mechanism.

The experimental data for $^{64}$Zn(n,α)$^{61}$Ni reaction cross section in the neutron energy range up to 15 MeV are very poor and there are significant discrepancies between the available results and estimations.

In this work we measured the cross sections and angular distributions for the $^{64}$Zn(n,α)$^{61}$Ni reaction at neutron energy $E_n$ =5.0, 5.7 and 6.5 MeV, using direct spectrometric method.

## II. Experimental Setup

The experiment was carried out at the 4.5 MV Van de Graaff Accelerator at the Institute of Heavy Ion Physics in Peking University. Nearly monoenergetic neutrons were produced via the $D(d,n)^3He$ reaction using a deuterium gas cell. The deuteron beam was ~3 μA during the experiment. **Figure 1** shows the scheme of the experimental arrangement.

Alpha-particles were detected using a parallel-plate, twin gridded ionization chamber with a common cathode. The chamber was made in the Frank Laboratory of Neutron Physics, JINR, Dubna, Russia.[1,2] Schematic view of the inner construction of the chamber is shown in **Fig. 2**.

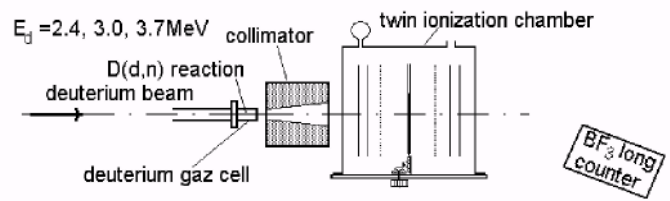

Fig. 1 Experimental arrangement

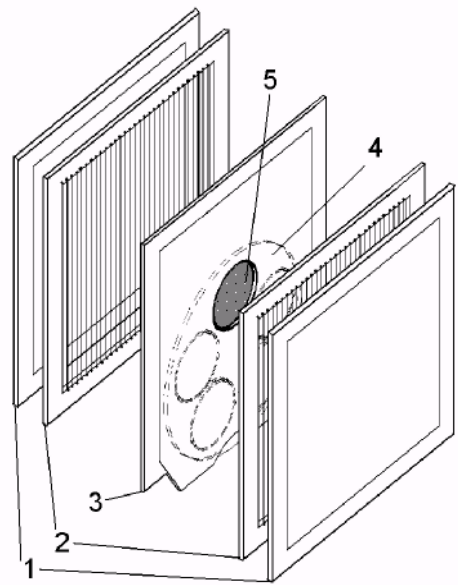

Fig. 2 Schematic view of the inner construction of the chamber.
1 - anode, 2 - Frisch grid, 3 - cathode, 4 - rotating disk, 5 - sample

* Corresponding author, Tel. +7-096-21-62113, Fax. +7- 096-21-65429, E-mail: gledenov@nf.jinr.ru

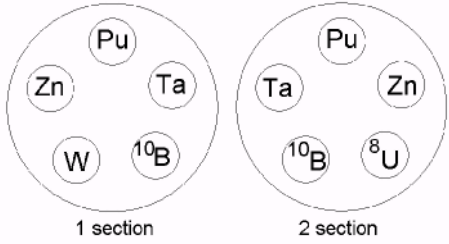

Fig. 3   Disposition of the targets on the cathode

Samples of Zn, an α-source, a background plate, $^{238}U$ were placed on the rotating disk of the cathode, as shown in Fig. 3. The thickness of the Zn targets was 266 μg/cm$^2$, the abundance of $^{64}Zn$ was 99.4% . The sample of Zn was prepared by evaporation in vacuum onto tantalum foil. The $^{238}U$ sample (total weight 7.85±0.1 mg, ϕ 4.5 cm) placed in the chamber was used to determine the absolute neutron flux.

The distances between the cathode and the grid, the grid and the anode were 4.5 cm and 2.2 cm, respectively. The counting gas was Kr+4.71% CH$_4$. The pressure of the counting gas was 1.2, 1.4, 1.6 atm for $E_n$=5, 5.7 and 6.5 MeV, respectively. The distance from the cathode to the center of the neutron source was ~38 cm.

The measurements were carried out from both sections simultaneously. The signals from the anode and the cathode were registered in coincidence and collected by multiparameter data acquisition system. This method provides us with the information about the energy and angular distribution of the emitted particles. At each neutron energy the sequence of measurements was performed: a main run with Zn, a background measurement run, and an $^{238}U$ run. During the experiment the neutron flux was monitored by the $BF_3$ long counter.

## III.   Experimental Results

**Figures 4 and** 5 show the two-dimensional spectra of α-particles for the $^{64}Zn(n,\alpha)^{61}Ni$ reaction at $E_n$ = 5.0 ± 0.26 MeV in the forward and backward directions.

Two α-lines from investigated reaction are clearly seen. Due to the facts that the daughter nucleus $^{61}Ni$ has high density of excited states, and the chamber has an energy resolution about 150 keV, these lines can be attributed to the transitions to ground and the first excited levels (the upper line in **Figs. 4, 5**) and to the group of levels from 4 to 7 (the lower line). The total cross section was determined on the sum of events (after the background subtraction) over the region limited by the Kr α-line.

**Table 1** Experimental cross sections of the $^{64}Zn(n,\alpha)^{61}Ni$ reaction.

| $E_n$ (MeV) | σ (mb) |
|---|---|
| 5.0±0.26 | 72.5±7 |
| 5.7±0.15 | 72.0±7 |
| 6.5±0.2 | 70.8±7 |

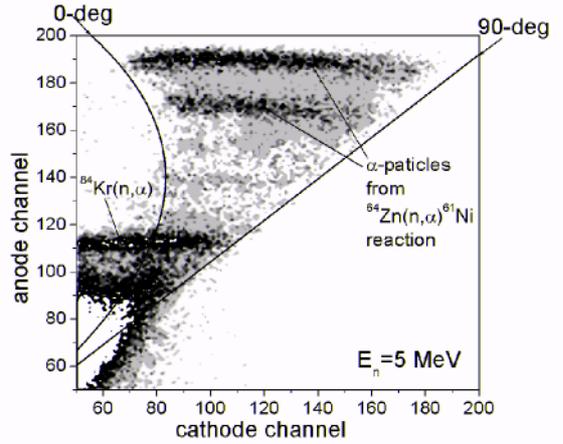

Fig. 4 Two-dimensional spectrum of $^{64}Zn(n,\alpha)^{61}Ni$ reaction at $E_n$=5 MeV in the forward direction

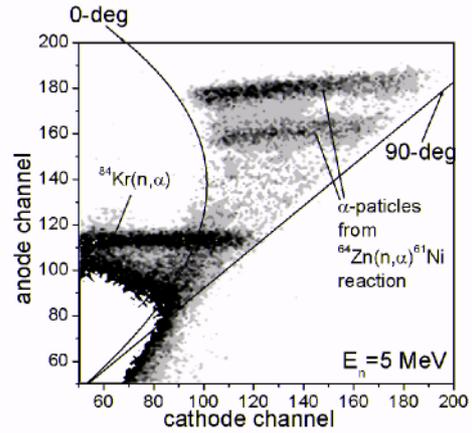

Fig. 5 Two-dimensional spectrum of $^{64}Zn(n,\alpha)^{61}Ni$ reaction at $E_n$=5 MeV in the backward direction

The values of the cross sections are given in **Table 1**.

The results for the angular distributions of the α-particles, emitted in the $^{64}Zn(n,\alpha)^{61}Ni$ reaction, are shown in the center-of-mass system in **Fig. 6**. The fitted curves are the Legendre polynomial presented as 1 + a · x +(b/2) · (3x$^2$ - 1) with parameters cited in **Table 2**.

**Table 2**   Parameters of angular distributions

| $E_n$ (MeV) | a | b |
|---|---|---|
| 5.0±0.26 | -0.019±0.016 | 0.393±0.028 |
| 5.7±0.15 | 0.163±0.029 | 0.299±0.045 |
| 6.5±0.2 | 0.212±0.034 | 0.562±0.051 |

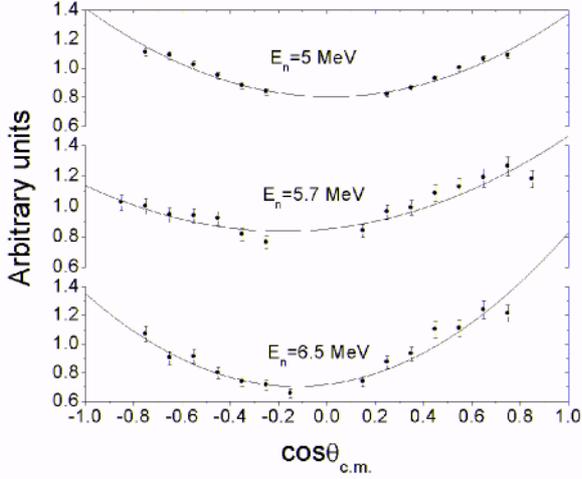

Fig. 6 Angular distributions for the $^{64}Zn(n,\alpha)^{64}Ni$ reaction

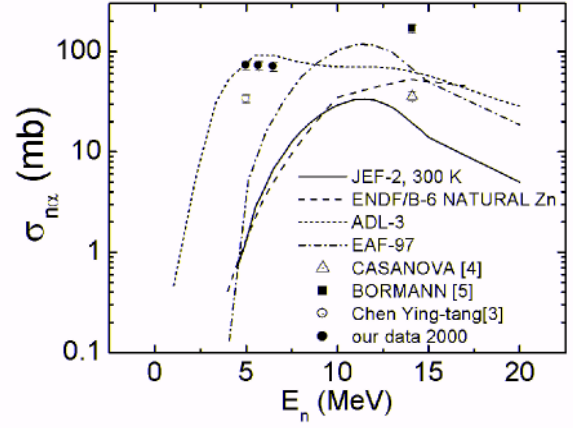

Fig. 7 Excitation function of the $^{64}Zn(n,\alpha)^{64}Ni$ reaction

It should be noted, that due to some technical reasons, normalization of the forward-backward directions of the angular distribution at 6.5 MeV was done using the Kr line and is not completely reliable. All the other indicated errors arose from statistics uncertainties of the measurements with Zn, background and $^{238}U$ measurements.

## IV. Discussion

Our results on the $^{64}Zn(n,\alpha)^{64}Ni$ cross section in comparison with other experimental data and theoretical evaluations are illustrated in **Fig. 7**. In neutron energy region of 1-10 MeV, the unique experimental data[3] has been also obtained in our preliminary measurements. As the first experiment suffered from a set of systematic uncertainties, this result could be considered as preliminary, and, in principal, is not in contradiction with the present data. Meantime, most of available evaluations are much less than the obtained experimental values. To verify this situation, we performed a theoretical estimation of the $^{64}Zn(n,\alpha)^{64}Ni$ cross section within this neutron energy region.

The cross section of the $^{64}Zn(n,\alpha)^{64}Ni$ reaction was determined using the method of the averaged cross section.[7,8] The proposed $(n,\alpha)$ cross section is:

$$\overline{\sigma_\alpha} = \frac{\lambda^2}{2} \sum_{I,J} \frac{2J+1}{2(2I+1)} \cdot \frac{\varepsilon_J^{Il}\langle \frac{\Gamma_n(l)}{D}\rangle \left(\frac{\Gamma_\alpha^l}{D_J^l}\right)}{\varepsilon_J^{Il}\langle \frac{\Gamma_n(l)}{D}\rangle + \left(\frac{\Gamma_\alpha^l}{D_J^l}\right)} \times$$
$$\times F\left(\frac{\frac{\Gamma_\alpha^l}{D_J^l}}{2\varepsilon_J^{Il}\langle \frac{\Gamma_n(l)}{D}\rangle}\right) \quad (1)$$

where I is the spin of the target, J is the spin of the compound nucleus, l is the the orbital momentum of the neutron.

$$\varepsilon_J^{Il} = \begin{bmatrix} 2, & |J-I| \le l \pm \frac{1}{2} \le |J+I| \\ 1, & |J-I| \le l \pm \frac{1}{2} \text{ or } l + \frac{1}{2} \le |J+I| \\ 0, & \text{in the rest} \end{bmatrix} \quad (2)$$

$$F(a) = (1+2a)\{1 - \sqrt{\pi a}\exp(a)[1 - Erf\sqrt{a}]\} \quad (3)$$

The neutron averaged parameters were taken from.[9] For the calculation of the $\alpha$ particle penetrabilities a Wood-Saxon potential[10] was used. The potential parameters are: $V_0 = -225$ MeV, $\alpha_r = 0.68$ fm, $r_r = 1.17$ fm.

For the first approach, we restricted the calculation by the s- and p-neutron waves and the α-transitions up to the 7-th excited level of $^{61}Ni$. The obtained results are shown in **Table 3.**

Table 3 Calculated cross sections of the $^{64}Zn(n,\alpha)^{64}Ni$ reaction

| $E_n$ (MeV) | $\sigma$ (mb) |
|---|---|
| 1 | 0.029 |
| 3 | 6.3 |
| 5 | 160 |

Our theoretical evaluation is two times greater than the experimental one at the $E_n = 5$ MeV incident neutron energy. There are several reasons explaining it. First, the potential parameters were chosen without any preliminary selection or fitting to the existing experimental data (for example, the elastic α-scattering). Second, and that is more important, we did not take into account the pre-equilibrium and direct processes. From **Fig. 6** and **Table 2**, it can be seen that there is a significant asymmetry of angular distributions at 5.7 and 6.5 MeV It could indicate to the contribution of direct processes. Obviously, the compound nuclear mechanism predominates at the energy lower 5 MeV, and over 5 MeV the contribution of direct process becomes significant. Thus, the ADL-3 evaluation curve is quite close to our data and probably has more realistic behaviour. An accurate evaluation of the cross section in the $^{64}Zn(n,\alpha)^{64}Ni$ reaction requires the knowledge of the parameters in the frame of a proposed model. These parameters could be obtained by fitting the experimental data. In our case, such an approach needs more experimental data in a larger neutron energy interval.

Analysing the existing experimental data from the literature on $(n,\alpha)$ reaction, with the increase of the neutron energy the contribution of the direct reaction mechanism in the cross section is increasing too (in comparison with the compound nucleus mechanism reaction). These main suppositions are at present under investigation.

**Acknowledgment**

We are gratefull to Prof. Yu.P.Popov of Frank Laboratory of Neutron Physics, JINR for fruitfull discussion on the theoretical and experimental aspects presented in this paper.

This research was supported partly by Russian Foundation for Basic Research, Grant N 98-02-17078.